\documentclass[
twocolumn,
showpacs,preprintnumbers,
 amsmath,amssymb,
 aps,pre,
]{revtex4-1}
\usepackage[english]{babel}
\usepackage[T1]{fontenc}
\usepackage[utf8]{inputenc}

\usepackage{graphicx}
\usepackage{hyperref}

\newcommand{\Ls}{L_\perp}
\newcommand{\Lp}{L_\parallel}
\newcommand{\Fc}{\mathcal{F}_\mathrm{C}}
\newcommand{\Fres}{\delta\hspace{-1pt}F}
\newcommand{\Dc}{\Delta_\mathrm{C}}
\newcommand{\Padd}{P_\mathrm{add}}
\newcommand{\Uc}{\delta\hspace{-1pt}f}
\newcommand{\Uctot}{\Fres}
\newcommand{\Fsc}{\Phi_\mathrm{c}}
\newcommand{\Tc}{T_\mathrm{c}}
\newcommand{\kB}{k_{\mathrm{B}}}
\newcommand{\ord}{\mathrm{o}}
\newcommand{\kme}{\kappa-1}
\newcommand{\Reff}{R_{\mathrm{eff}}}

\newcommand{\etal}{\textit{et~al.~}}
\newcommand{\eg}{e.\,g.}
\newcommand{\ie}{i.\,e.}

\usepackage{color}
\newcommand{\fred}[1]{{\color[rgb]{0.9,0.3,0.3}#1}}

\let \vec=\mathbf

\newcommand{\Del}{\Delta}
\newcommand{\rdrei}{r_3}
\newcommand{\zmin}{z_\mathrm{min}}
\newcommand{\zmax}{z_\mathrm{max}}
\newcommand{\expo}{q}

\begin{document}

\title{Many-body critical Casimir interactions in colloidal suspensions}

\author{Hendrik Hobrecht}
\author{Alfred Hucht}%

\affiliation{Fakultät für Physik and CENIDE, Universität Duisburg-Essen, D-47048 Duisburg, Germany}

\date{October 20, 2015}

\begin{abstract}
We study the fluctuation-induced Casimir interactions in colloidal suspensions, especially between colloids immersed in a binary liquid close to its critical demixing point.
To simulate these systems, we present a highly efficient cluster Monte Carlo algorithm based on geometric symmetries of the Hamiltonian.
Utilizing the principle of universality, the medium is represented by an Ising system while the colloids are areas of spins with fixed orientation.
Our results for the Casimir interaction potential between two particles at the critical point in two dimensions perfectly agree with the exact predictions.
However, we find that in finite systems the behavior strongly depends on whether the $Z_{2}$ symmetry of the system is broken by the particles.
Eventually we present Monte Carlo results for the three-body Casimir interaction potential and take a close look onto the case of one particle in the vicinity of two adjacent particles, which can be calculated from the two-particle interaction by a conformal mapping.
These results emphasize the failure of the common decomposition approach for many-particle critical Casimir interactions.
\end{abstract}

\maketitle

\section{Introduction}

Over the last few years colloidal suspensions have played a key role as experimental model systems for phase transitions and aggregation processes.
Especially colloids immersed in a binary liquid have been investigated experimentally, since near the demixing transition of the solvent long-ranged correlated fluctuations give rise to critical Casimir forces.
Those forces are peculiarly interesting for several qualities:
Experimentally their sensitive temperature dependency allows to control the interaction strength \textit{in situ} and reversibly.
Furthermore the colloids can be observed directly with common microscopy methods.
From a theoretical point of view, these critical Casimir forces are interesting because they are universal, \ie, different systems in the same universality class show the very same behavior and share universal scaling functions near their critical point.

Fluctuation-induced forces were first predicted by Casimir in 1948 \cite{Casimir48}; he realized that the fluctuations of the electromagnetic field cause an attractive force between two perfectly conducting plates in vacuum.
Analogously, there are critical Casimir forces near a continuous phase transition in thermodynamic systems induced by the long-ranged correlated fluctuations of the order parameter if the medium is restricted by the systems geometry.
An according theory was formulated in 1978 by Fisher and de Gennes for a system in $d$-dimensional slab geometry $\Ls\times \Lp^{d-1}$ with $\Lp\gg\Ls$ \cite{FisherdeGennes78}.
A first experimental proof of those forces has been given by Garcia and Chan; they observed a change in the thickness of critical liquid films of $^{4}$He near its $\lambda$-point \cite{GarciaChan99} and of $^{3}$He-$^{4}$He mixture near its tricritical point \cite{GarciaChan02}.
Binary liquids have also been studied in this geometry, extending the experiments to the Ising universality class \cite{FukutoYanoPershan05}.
Hertlein \etal reported the direct measurement of the critical Casimir force between a single colloid and a wall embedded in a binary liquid \cite{HertleinHeldenGambassiDietrichBechinger08}, which led to several other experiments on colloidal suspensions \cite{SoykaZvyaHertHeldBech08,BonnOtwiSacaGuoWegSchall09,ZAB11,DVNBS13}.

Since there are only a few simple systems for which the scaling functions are known exactly, \eg, the two-dimensional Ising model \cite{EvansStecki94,AbrahamMaciolek10,HuchtGruenebergSchmidt11} or the large-$n$ approach \cite{DieGruHaHuRuSch12,DieGruHaHuRuSch14a}, both in slab geometry, Monte Carlo (MC) studies of such critical systems have proven to be another tool to determine the characteristic universal behavior.
Besides the MC simulations of the slab geometry, which all use some kind of thermodynamic integration to determine the scaling functions of the free energy and the Casimir force itself \cite{Hucht07a,VasilyevGambassiMaciolekDietrich09}, there are a few studies recent on the interaction between a spherical particle and a wall \cite{Hasenbusch13,HobrechtHucht14} and between fixed spherical particles \cite{ZMD13,Vasilyev14}.
In a recent study we have shown how the thermodynamic integration can be avoided by allowing an object to move during the MC simulation and analyzing its distribution function \cite{HobrechtHucht14}. 
This approach is therefore very similar to the experimental method of Hertlein \cite{HertleinHeldenGambassiDietrichBechinger08}, and it was used very recently to study the phase diagram of two-dimensional colloids \cite{ETBEvRD15}.

As the Casimir forces are nonadditive, describing their interactions with pair potentials is insufficient in systems with more than two particles, and higher-order contributions have to be taken into account, like done in \cite{MattosHarnauDietrich2015} within a mean-field approach for cylindrical colloids.
Unfortunately the common methods to simulate such systems with many-body interactions get inefficient or inaccurate the more bodies interact. 
Previous simulation studies \cite{DVNBS13, HobrechtHucht14, ETBEvRD15} used standard local Monte Carlo algorithms, which suffer from the time-consuming effect of critical slowing down near the critical point and thus become very inefficient especially for large systems.
This effect can be suppressed by using cluster algorithms; the probably most famous example is the Wolff algorithm \cite{Wolff89}, first introduced for the Ising model and based on the previous work by Swendsen and Wang \cite{Swendsen87}.
While those algorithms utilize the $Z_{2}$ symmetry of the Hamiltonian and do not conserve the order parameter, Heringa and Bl{\"o}te introduced an algorithm which uses the invariance of the Hamiltonian under some geometric transformations, \eg, a point reflection \cite{HeringaBl98}, builds two symmetric clusters and exchanges them under conservation of the order parameter.
This geometric cluster algorithm (GCA) was later generalized to an off-lattice algorithm for spherical particles by Liu and Luijten, which builds two clusters of particles and exchanges them \cite{LiuLuijten04}.
Therefor this generalized geometric cluster algorithm reflects a particle at a pivot point and iteratively adds spheres to the cluster utilizing the hard-sphere potential, until a configuration without overlaps is generated.
The cluster building process can be extended to include additional interactions between the particles.

In the present study we introduce a new MC cluster algorithm that is capable of both moving the particles and mixing the medium within one cluster step.
We apply this algorithm to the two-dimensional Ising model in order to compare our results to the exactly known universal two-body interaction at criticality, calculated by Burkhardt and Eisenriegler \cite{BurkhardtEisenriegler95} using conformal field theory \cite{Cardy84}.
Therefor we will introduce the common theoretical background, starting with the description of the critical Casimir effect in slab geometry and its connection to the fluctuation-induced force between two colloidal particles in a critical medium.
Having recapitulated the exact results for the interaction potential scaling function, we explain our method to determine the potential itself using the two-particle correlation function.
Afterwards we present the model we use to simulate colloidal suspensions.
We describe the cluster algorithm and discuss the crucial modifications compared to the original GCA as well as its limiting cases.
Eventually we compare our MC results for two interacting particles with the theory and discuss the various non-negligible finite-size effect that occur if the symmetry of the medium is broken by the particles and how those effects can be handled.

Coming from the exact results for the two-particle interaction, we develop a conformal mapping from the known case of the annulus geometry onto a three-body setting with two adjacent disks touching at their closest approach and a third, slightly distorted one free to move around them.
From this mapping, we calculate an approximation of the according three-body Casimir interaction scaling function and its asymptotical amplitude for large distances.

Afterwards we present first results for the three-body interactions.
Therefore we show how the $n$-body correlation function and the $n$-body Casimir interaction potential are connected in general.
We introduce the concept of an infinitely strong \textit{ghost bond} between two particles acting as constraint to the possible particle motions. 
With this ghost bond we are able to considerably speed up the simulations.
We present our MC results for the three body interaction at criticality, which emphasize that the common decomposition into two-particle interactions and -- in comparison weak -- higher-order contributions is insufficient, since the pure three-body interaction is of the same order as the one for two-particles.
To validate these results we compare the simulation results with the approximation from the conformal mapping.
Finally we comment on the consequences of these results for particle clusters interacting with Casimir forces.

\section{Theory}

Before we head for the critical Casimir interaction potential between two spherical objects, we start with a short repetition of the relevant quantities in the much simpler slab geometry $\Ls\times\Lp^{d-1}$ with periodic boundary conditions (BC) in the parallel directions and arbitrary BCs in perpendicular direction.
The critical Casimir force per unit area $A=\Lp^{d-1}$ at reduced temperature $t=T/\Tc-1$ reads
\begin{align}
  \beta\Fc (t,\Ls,\Lp)\equiv -\frac{1}{A}\frac{\partial}{\partial \Ls}\Fres(t,\Ls,\Lp),
\end{align}
with $\beta=1/\kB T$, where the residual free energy $\Fres$ -- also called \textit{Casimir potential} in the context of colloidal suspensions -- is given by
\begin{align}
  \Fres(t,\Ls,\Lp)\equiv F(t,\Ls,\Lp)-V\!f_{\mathrm{b}}(t)-A f_{\mathrm{s}}(t),
\end{align}
with total free energy $F(t,\Ls,\Lp)$, bulk free energy density $f_{\mathrm{b}}(t)$, surface free energy per unit area $f_{\mathrm{s}}(t)$, and volume $V=\Ls A$. 
All energies are in units of $\kB T$.

Fisher and de Gennes \cite{FisherdeGennes78} proposed that the Casimir force fulfills the scaling \textit{ansatz}
\footnote{Throughout this work, the symbol $\simeq$ means asymptotically equal in the respective limit, e.g., $f(L)\simeq g(L)\Leftrightarrow\lim_{L\rightarrow\infty}f(L)/g(L)=1$.}
\begin{align}
  \beta\Fc(t,\Ls,\Lp)\simeq\Ls^{-d}\vartheta^{ab}(x,\rho)
\end{align}
near the critical point, where $a,b \in \{\uparrow,\downarrow,\ord\}$ denote the surface preferences of the boundaries.
Beneath those surface preferences, the universal scaling function $\vartheta^{ab}$ depends only on the temperature scaling variable $x$ and on the given geometry, represented by the aspect ratio $\rho$,
\begin{align}
  \label{eq:Def_x_rho}
  x=t\left(\frac{\Ls}{\xi_{0}^{+}}\right)^{1/\nu},\quad\rho=\frac{\Ls}{\Lp},
\end{align}
where $\xi_{0}^{+}$ is the correlation length amplitude above $\Tc$, and $\nu$ is the critical exponent of the correlation length.
Accordingly, the Casimir potential satisfies a similar \textit{ansatz}
\begin{align}
  \Fres(t,\Ls,\Lp)\simeq\Phi^{ab}(x,\rho),
\end{align}
where $\Phi^{ab}$ is again a universal scaling function.

Using this result for $\rho\to 0$ and the conformal invariance at criticality \cite{Cardy84}, Burkhardt and Eisenriegler calculated the asymptotics of the Casimir potential for two spherical objects in an infinitely large $d$-dimensional critical medium at distances small compared to their radii \cite{BurkhardtEisenriegler95}.
Analogously to the geometric scaling variable $\rho$ in Eq.~\eqref{eq:Def_x_rho}, they introduced the conformal invariant scaling variable
\begin{align}
  \label{eq:kappa}
  \kappa=\frac{r^{2}-R_{1}^{2}-R_{2}^{2}}{2R_{1}R_{2}},
\end{align} 
where $R_{1}$ and $R_{2}$ are the radii of the two particles with surface preferences $a$ and $b$, respectively, and $r$ is the distance between their centers.
Thus the conformal scaling variable $\kappa$ encodes both the relative positions and the sizes of the two spheres.

For arbitrary $d$ the Casimir potential scaling function of two spheres at the critical point fulfills
\begin{align}
  \label{eq:ScalingSpheres}
  \Fsc^{ab}(\kappa)\simeq\Fres_2(t{=}0,R_{1},R_{2};r),
\end{align}
where $\Fres_2$ is the free energy of the system with two particles at distance $r$, and asymptotically reads ($\Fsc^{ab}$ is named $\mathcal F_{ab}$ in \cite{BurkhardtEisenriegler95})
\begin{align}
  \label{eq:FscKappa}
  \Fsc^{ab}(\kappa)\stackrel{\kappa\to 1^+}{\simeq}\Dc^{ab} \, S_{d}\,[2(\kme)]^{-(d-1)/2}
\end{align}
for small distances $\kappa\to 1^+$, where $S_{d}$ is the surface area of the $d$-dimensional unit sphere and $\Dc^{ab}$ denotes the universal Casimir amplitude of the corresponding slab geometry. 
For the two-dimensional Ising case considered in this work $S_2=2\pi$, and $\Dc^{\uparrow\uparrow}=\Dc^{\ord\ord}=-\pi/48$, as well as $\Dc^{\uparrow\downarrow}=23\pi/48$.

In the limit of large distances between the two particles, \ie, for $\kappa\gg 1$, the Casimir potential is dominated by the correlation functions of the relevant operators $\langle\phi\phi\rangle$, as can be shown by a \textit{small sphere expansion} of the Boltzmann factor analogously to the common operator product expansion of a conformal field theory \cite{BurkhardtEisenriegler95}, and thus behaves as
\begin{align}
  \label{eq:Fsc_large}
  \Fsc^{ab}(\kappa)\stackrel{\kappa\gg 1}{\simeq}-\sum_{\phi}Q^{ab}_{\phi}[2(\kappa+1)]^{-x_{\phi}},
\end{align}
where the sum is over all relevant scaling operators $\phi$, \eg, for the Ising model the energy density $\epsilon$ and the magnetization density $\sigma$, with according scaling dimension $x_{\phi}$.
The amplitude ratios $Q^{ab}_{\phi}$ are universal and are known exactly for some special cases, \eg, the two-dimensional Ising class \cite{BurkhardtEisenriegler95}, where Eq.~\eqref{eq:Fsc_large} simplifies to
\begin{subequations}
\label{eq:asymp2}
\begin{align}
  \Fsc^{\uparrow\uparrow}(\kappa \gg 1)& \simeq -\sqrt{2}[2(\kappa+1)]^{-1/8},\\
  \Fsc^{\uparrow\downarrow}(\kappa \gg 1)& \simeq \sqrt{2}[2(\kappa+1)]^{-1/8},\\
  \Fsc^{\ord\ord}                (\kappa \gg 1)& \simeq -           [2(\kappa+1)]^{-1}.
\end{align}
\end{subequations}
For completeness we also give the exact scaling functions for the two-dimensional Ising case \cite{Cardy2006, BimonteEmigKardar13},
\begin{align}
  \label{eq:exact}
  \Fsc^{ab}(\kappa)& = \frac{\pi \rho}{12}-\ln\!\left[
		\sqrt{\frac{\vartheta_3(e^{-2\pi\rho})}{\eta(2 i \rho)}}
   + s_{ab}	\sqrt{\frac{\vartheta_2(e^{-2\pi\rho})}{\eta(2 i \rho)}}\right].
\end{align}
Here, $\kappa=\cosh(2\pi\rho)$ and $s_{ab} = \{1,0,-1\}$ for BCs $ab=\{\uparrow\uparrow, \ord\ord, \uparrow\downarrow\}$, while $\vartheta_j(q)$ and $\eta(\tau)$ denote the Jacobi theta functions and the Dedekind eta function, respectively.
Indeed, $\rho$ is the aspect ratio of the corresponding system in slab geometry with periodic BCs in one direction, i.e., the surface of a cylinder, which can be conformally mapped onto two separated disks.
Note that for the two-dimensional case the interaction between any two arbitrary shaped objects may be obtained exactly \cite{BimonteEmigKardar13}, as the group of conformal transformations is more powerful in $d=2$.

\begin{figure}[b]
  \includegraphics[width=0.667\columnwidth]{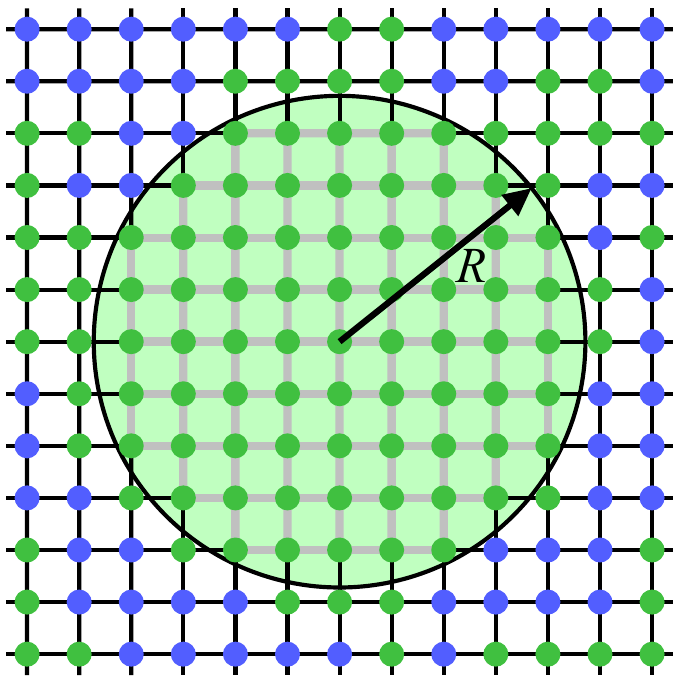}
  \caption{(Color online) Embedment of a spherical particle with radius $R$ onto the lattice. 
  $\uparrow$ and $\downarrow$ spins are green (light) and blue (dark), respectively. 
  Infinite strong couplings are grey, while normal bonds with strength $K$ are shown black.
  }
  \label{fig:system}
\end{figure}

The reversible work theorem \cite{chandler1987introduction} states that the residual free energy $\Fres_2(r)$ is directly related to the two-particle distribution function $g_2(r)$ by
\begin{align}
  \label{eq:ReversibleWork2}
  g_2(t,R_{1},R_{2};r)=\mathrm e^{-\Fres_2(t,R_{1},R_{2};r)}.
\end{align}
Note that in the bulk $\Fres_2(r{\to}\infty)=0$ and $g_2(r{\to}\infty)=1$ by definition.
If the system is at its critical point, we can use Eqs.~\eqref{eq:ScalingSpheres} and \eqref{eq:ReversibleWork2} to calculate the corresponding Casimir potential scaling function 
\begin{align}
  \label{eq:ReversibleWork3}
  \Fsc^{ab}(\kappa)\simeq-\ln g_2(t{=}0,R_{1},R_{2};r),
\end{align}
and hence we can determine $\Fsc^{ab}(\kappa)$ from a measurement of the two-particle distribution function $g_2(r)$. 
The results of this analysis are presented in Section~\ref{sec:twopoint}.

\section{Model and Cluster Algorithm}

\begin{figure*}[t]
  \includegraphics[width=0.24\textwidth]{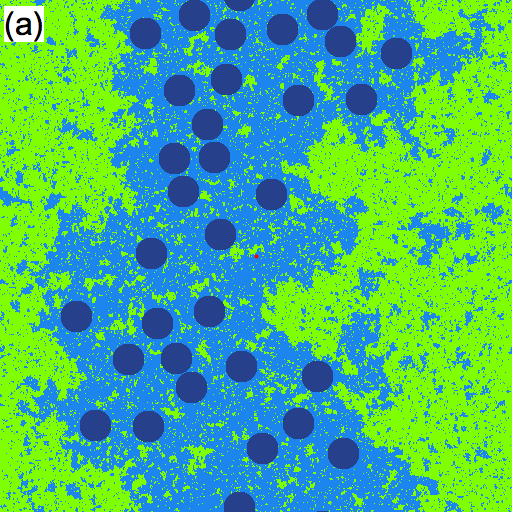}\hfill%
  \includegraphics[width=0.24\textwidth]{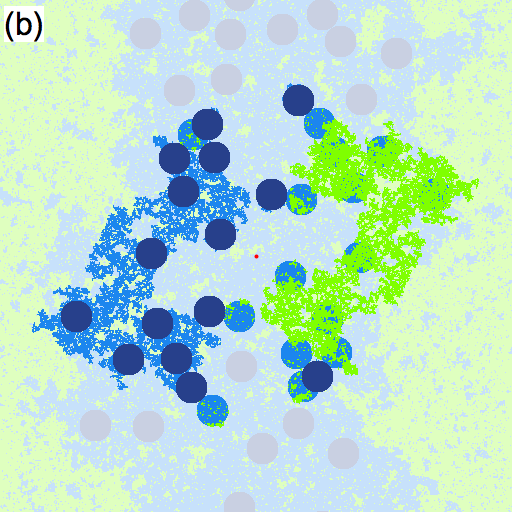}\hfill%
  \includegraphics[width=0.24\textwidth]{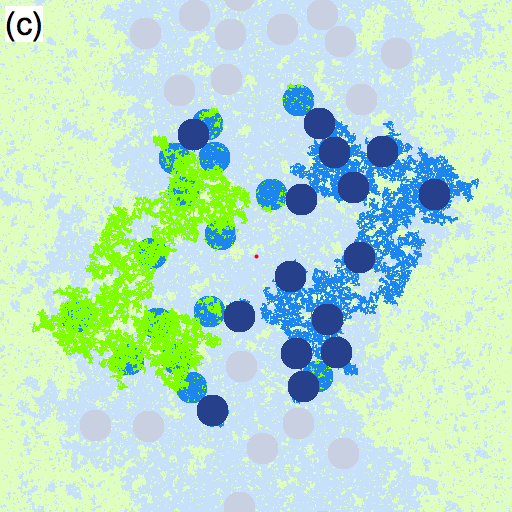}\hfill%
  \includegraphics[width=0.24\textwidth]{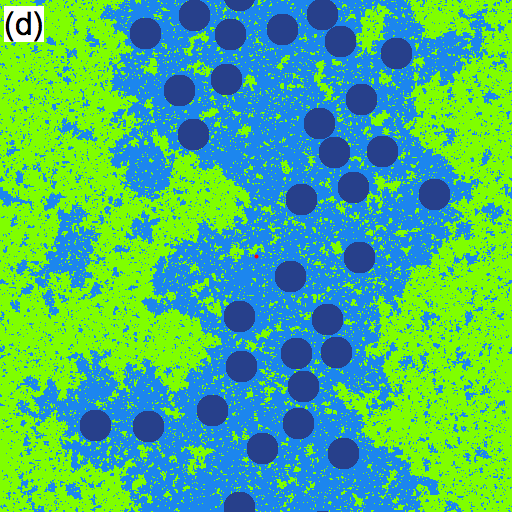}
  \caption{(Color online) One typical cluster spin flip in a periodic two-dimensional system with linear size $L=512$ and $N_{\uparrow}=32$ particles with radius $R=16$ and $\uparrow$ surface preference at $\Tc$. 
  Up (down) medium spins are shows in blue (light green). 
  The medium magnetization is fixed to $M=0$ and the pivot (red point) is in the center of the figures. 
  Starting from an initial configuration (a) two clusters are build symmetrically around the pivot (b) and are exchanged (c), leading to a new equilibrium configuration (d).}
  \label{fig:clusterspinflip}
\end{figure*}

We model the binary liquid using an Ising system with the Hamiltonian, in units of $\kB T$,
\begin{align}
  \mathcal{H}=-\sum_{\langle ij \rangle}K_{ij}s_{i}s_{j}
\end{align}
on a simple cubic lattice in $d$ dimensions with $L^d$ spins and periodic BCs in all directions.
The reduced couplings $K_{ij}\geq0$ are assumed to be ferromagnetic, and the sum runs over all nearest neighbor pairs $\langle ij \rangle$.
Analogous to a lattice-gas interpretation, the spins pointing up ($s\;{=}\uparrow$) or down ($s\;{=}\downarrow$) may be understood as particles of species $A$ or $B$, respectively.
The critical composition for this system is the ratio $A{:}B=1{:}1$, \ie, at total medium magnetization $M=0$.

Now we insert $N$ spherical particles with radius $R_\mu$ located at positions $\vec r_\mu$, $\mu=1,\ldots,N$, into the system, see Fig.~\ref{fig:system}. 
Each particle is realized as a group of spins at positions $\vec r$ fulfilling   $|\vec r - \vec r_\mu| < R_\mu$, aligned in the same direction by virtue of infinitely strong couplings according to
\begin{align}
\label{eq:Jij}
  K_{ij}=\begin{cases}
    \infty &\text{if }s_{i}\text{ and }s_{j}\text{ belong to one particle,}\\
    K&\text{else.}
    \end{cases}
\end{align}
For a more complex model one could also distinguish between particle-medium couplings, the coupling between two different particles at the particle surfaces, or even use locally varying couplings.

Apart from the nearest neighbor couplings $K_{ij}$ between the spins there are no other particle-particle interactions.
Instead, these interactions are induced by the correlated medium, and thus not only the Casimir pair interaction, but all fluctuation induced many-body interactions are included automatically in our method.

We now modify the GCA by Heringa and Bl{\"o}te and explicitly include the bonds $\langle ij \rangle$ with couplings $K_{ij}$ into the cluster building process. 
This way the particles encoded into the bonds will also be moved by the cluster algorithm.
We assume that neighboring lattice sites $i$ and $j$ as well as the connecting bond $\langle ij \rangle$ are mapped onto the sites $i'$ and $j'$ and the bond $\langle i'j' \rangle$, respectively.
We assume this mapping to be a point reflection with respect to a pivot as symmetry operation (although it could be realized by any geometric mapping that leaves the Hamiltonian invariant).

\begin{figure*}[t]
{}\hfill
  \includegraphics[height=0.19\textwidth]{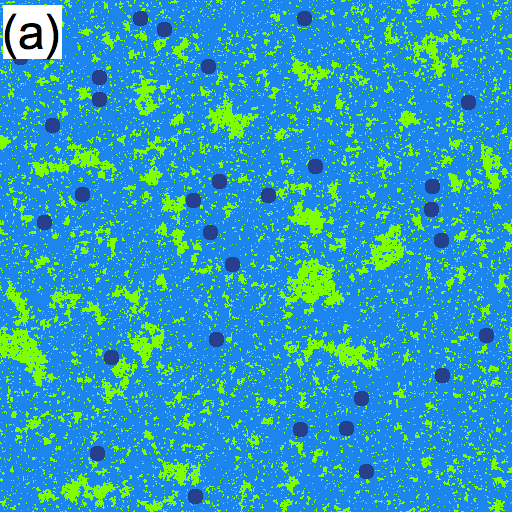}\hfill%
  \includegraphics[height=0.19\textwidth]{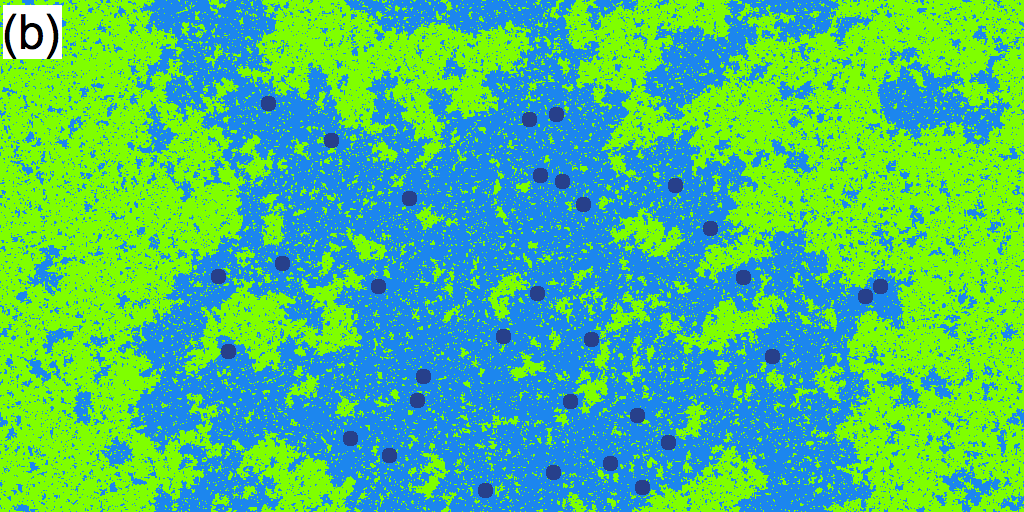}\hfill%
  \includegraphics[height=0.19\textwidth]{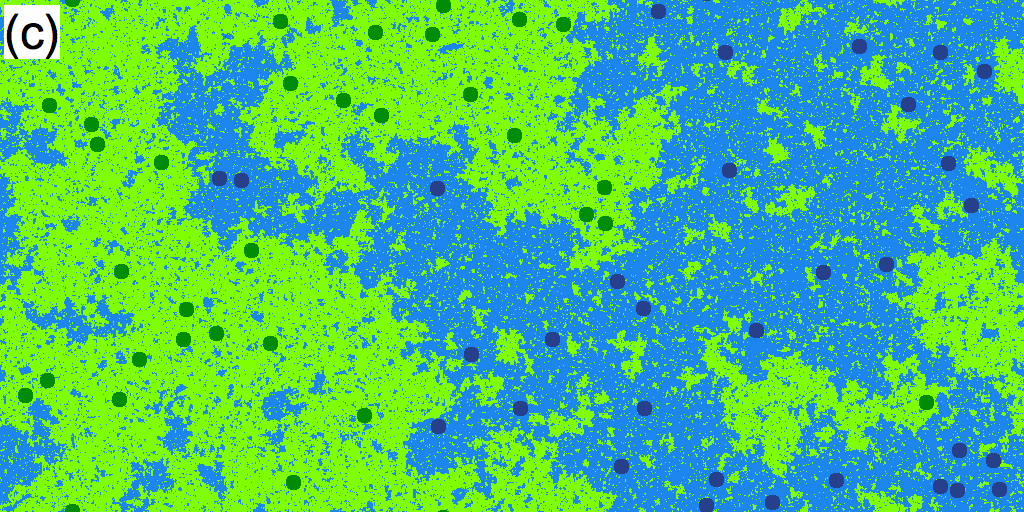}%
  \hfill{}
  \caption{(Color online) Typical configurations of three different periodic systems at $\Tc$ with particles having radius $R=8$. 
  (a) $L_{x}\times L_{y}=512\times512$ system with fluctuating magnetization and $N_{\uparrow}=32$ particles with $\uparrow$ surface preference. 
  The polarization leads to strong deviations from the critical behavior.
  (b) $L_{x}\times L_{y}=1024\times512$ system with fixed magnetization and $N_{\uparrow}=32$ particles with $\uparrow$ surface preference.
  The two dominant domains present in the system lead to a more realistic critical domain structure.
  (c) $L_{x}\times L_{y}=1024\times512$ system with fluctuating magnetization, $N_{\uparrow}=32$ particles, and $N_{\downarrow}=32$ particles with according surface preference.
  Due to the symmetric polarization effect, the system behaves like case (b).
  }
  \label{fig:configurations}
\end{figure*}

Starting with the spins $s_{i}$ and $s_{i'}$, the energy difference of an exchange of the two spins $s_{j}$ and $s_{j'}$ due to activation of the bond pair $\langle ij \rangle$ and $\langle i'j' \rangle$ is given by
\begin{align}
  \label{eq:DeltaE}
  \Delta E = K_{ij}s_{i}s_{j} + K_{i'j'}s_{i'}s_{j'} - K (s_{i}s_{j'} + s_{i'}s_{j}),
\end{align}
where the cases of infinite couplings, Eq.~\eqref{eq:Jij}, are accounted for correctly by using a finite coupling $K$ in the last term.
The algorithm can be summarized as follows:
{
\setlength{\leftmargini}{12pt}
\setlength{\leftmarginii}{20pt}
\begin{enumerate}\itemsep2pt \parskip0pt \parsep0pt
\item Randomly choose two different lattice sites $i$ and $i'$ and use them as starting points for the two clusters $\mathcal C$ and $\mathcal C'$. 
Set the pivot to the midpoint of $i$ and $i'$. 
Put $(i,i')$ on the stack (a list of lattice site pairs). 
\item Read a pair $(i,i')$ from the stack.
For all neighbor sites $(j,j')$ of $(i,i')$ calculate $\Delta E$ (Eq.~\eqref{eq:DeltaE}) and, if $\Delta E>0$, do the following with probability $\Padd=1-\mathrm e^{-\Delta E}$:
\begin{enumerate}\itemsep1pt \parskip0pt \parsep0pt
\item Add the bonds $\langle ij \rangle$ and $\langle i'j' \rangle$ to  $\mathcal C$ and $\mathcal C'$ if they are not already added.
\item If $(j,j')$ are not already added to the clusters, add them to $\mathcal C$ and $\mathcal C'$ and put them on the stack.
\end{enumerate}
Else, do nothing.
\item Execute step (2) until the stack is empty.
\item Exchange the clusters $\mathcal C$ and $\mathcal C'$.
\end{enumerate}
Note that step 4 can be eliminated, as the spin and bond exchange can already be performed during the cluster building process. 
Furthermore, it is sufficient to put spin $i$ on the stack and to calculate $i'$ using the pivot.
}

If both $s_i$ and $s_j$ or both $s_{i'}$ and $s_{j'}$ belong to one particle, the energy difference $\Delta E=\infty$ and thus $\Padd=1$, which ensures that particles are always added as a whole and that there are only bonds with coupling $K$ at the edges of the two clusters.
Thus the proof of detailed balance follows just the one for the GCA \cite{HeringaBl98} with the energy $\Delta E=K(s_{i}-s_{i'})(s_{j}-s_{j'})$ for activating an edge bond.
Note that the activated bonds within the clusters, especially all bonds of a particle, are exchanged, too, so the particles do not fall apart.

If no particles are initialized, this algorithm is identical to the original GCA, which has its percolation threshold at the critical point and thus suppresses the effect of critical slowing down very efficiently.
Additionally, there is a second limiting case: for infinite high temperature, i.e., $K=0$ in Eq.~\eqref{eq:Jij}, this algorithm is a lattice version of the algorithm by Liu and Luijten \cite{LiuLuijten04}, since for this case the medium is not correlated anymore and the particles do only interact with a hard-sphere potential.

It is straightforward to include walls as lines of infinitely coupled $\uparrow$ or $\downarrow$ spins into the algorithm similar to Ref.~\cite{HobrechtHucht14}. 
Furthermore, we can include arbitrary particle-particle couplings as long-ranged bonds with strength $K_{\mu\nu}(\vec r_\mu,\vec r_\nu)$ between the particle centers.

In the following sections we use this algorithm to investigate a two-dimensional system of size $L_{x}\times L_{y}$ with $N_{\uparrow}$ identical spherical particles with $\uparrow$ surface preference. We first turn to the two-particle interaction.

\section{Two-particle interaction}\label{sec:twopoint}

As the simulations are performed on a lattice, rotational invariance is broken and distribution functions like $g_2$ must be considered as functions of the distance vector $\vec r$ instead of the scalar distance $r$, where $\vec r=\vec r_\nu - \vec r_\mu$ denotes the discrete distance vector between the centers of two particles $\mu$ and $\nu$.
Thus we analyzed the two-particle distribution function $g_2(\vec r)$ of the particles at criticality and at sufficiently low particle volume fractions $\varrho\equiv N\pi R^{2}/(L_{x}L_{y})$, obtained the Casimir potential $\Fres_2(\vec r)$ using Eq.~\eqref{eq:ReversibleWork2} and compared it with the exact scaling function $\Fsc^{ab}(\kappa)$ for $\varrho\to 0$.

To determine the Casimir potential we made a histogram of the particle distances $\vec r$, based on at least 6 million independent particle positions. 
Normalization of this histogram directly gives the discrete two-dimensional two-particle distribution function $g_2(\vec r)$.
Then we assigned the appropriate value of the conformal invariant scaling variable $\kappa=|\vec r|^{2}/(2 \Reff)-1$, Eq.~\eqref{eq:kappa}, to each point of $g_2(\vec r)$, where the leading lattice discretization effects are corrected with an effective radius $\Reff=R+\delta R$ for the disks, with $\delta R=-0.8(1)$, as proposed in Refs.~\cite{Hasenbusch13,HobrechtHucht14}.
Additionally, we used a logarithmic binning in $\kme$ for the data to obtain equidistant points in the double logarithmic scale and for the sake of a better statistic.
With the MC data we were able to cover a range of approximately $\kme\in[10^{-2},10^{3}]$.
Having obtained the two-particle distribution function, we finally used Eq.~\eqref{eq:ReversibleWork2} to determine the interaction potential $\Fres_2(\vec r)$.

\begin{figure}[t]
  \includegraphics[width=0.48\textwidth]{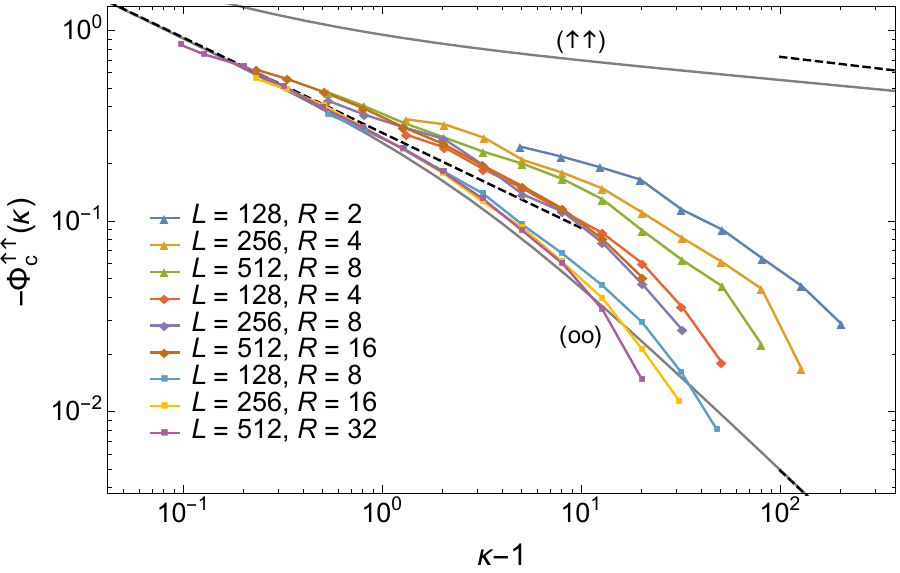}%
  \caption{(Color online) Two-particle Casimir potential $\Fsc^{\uparrow\uparrow}(\kappa)$ between $N_{\uparrow}$ $\uparrow$-spheres with radius $R$ embedded into a medium of size $L^2$ with fluctuating order parameter at $T=\Tc$.
The data points are $-\Fres_2(\vec r)$ from the simulations, Eq.~\eqref{eq:ReversibleWork2}, while the solid lines are calculated using Eq.~\eqref{eq:exact}, and the dashed lines are the asymptotes from Eqs.~\eqref{eq:FscKappa} and \eqref{eq:asymp2}.
 The MC results do not fall onto the predicted curve for $\uparrow\uparrow$ BCs.
 Due to the finite system size $L$ the spheres polarize the medium and the system is shifted off criticality.
 However, the data for fixed volume fraction $\varrho$ (marked with the same symbols) collapse onto each other for $R\geq8$, while smaller disks deviate upwards. 
 }
 \label{fig:Fab1}
\end{figure}

As it is common to study systems with a fluctuating instead of a fixed order parameter, we combined our cluster algorithm with Wolff cluster updates for the spin medium.
We simulated particles with $R\in\{2,4,8,16,32\}$ embedded in square systems with $L=L_{x}=L_{y}\in\{128,256,512\}$, using three different combinations for each value of the volume fraction $\varrho$.
In this context one would expect the determined potentials $\Fres_2(\vec r)$ to fall on the predicted curves for the symmetry-breaking surface preferences $\Fsc^{\uparrow\uparrow}(\kappa)$ as shown by Machta \textit{et al.} \cite{Machta2012}, but we find them to continuously interpolate between the expected behavior and the scaling function $\Fsc^{\ord\ord}(\kappa)$ with an additional exponential cut-off at about $r\approx L/2$ as shown in Fig.~\ref{fig:Fab1}.
However, systems with the same volume fraction $\varrho$ collapse nicely onto each other for $R\geq8$, while the systems with smaller disks deviate from this behavior.
This kind of behavior can be understood: The particles act as surface magnetic fields, forcing the surrounding medium to form a domain with the same orientation, which is described by the decay of magnetization profile around such a particle proportional to $r^{-1/4}$ \cite{BurkhardtEisenriegler85}.
Due to this slow decay and the periodic boundary conditions in both directions, finite-size effects of the order of $L^{-\beta/\nu}$, with $\beta=1/8$ and $\nu=1$, occur for the medium magnetization.
The system size is way too small to show bulk behavior, and the whole system is forced to polarize as illustrated in Fig.~\ref{fig:configurations}a.
This polarization effect shifts the system away from the critical point and, since the large distance behavior is dominated by the magnetization and energy correlation functions (Eq.~\eqref{eq:asymp2}), leads to the characteristic exponential cut-off.
Additionally the influence of the magnetization correlation decreases as the mean magnetization $M$ grows with the particle volume fraction $\varrho$. 
Thus the free energy is dominated by the energy correlation function, which is the only relevant operator for open boundary conditions, and it seems like the free energy converges against the according scaling function.
But this is not the limiting case for large $\varrho$; indeed at some value a density induced cluster process will start and the analysis for small $\varrho$ does not hold true anymore.
For small volume fraction $\varrho$ the polarization effect becomes smaller, thus the effect should vanishes in the limit $L\to\infty$ with $R$ fixed. 

\begin{figure*}[t]
	{}\hfill%
  	\includegraphics[width=0.48\textwidth]{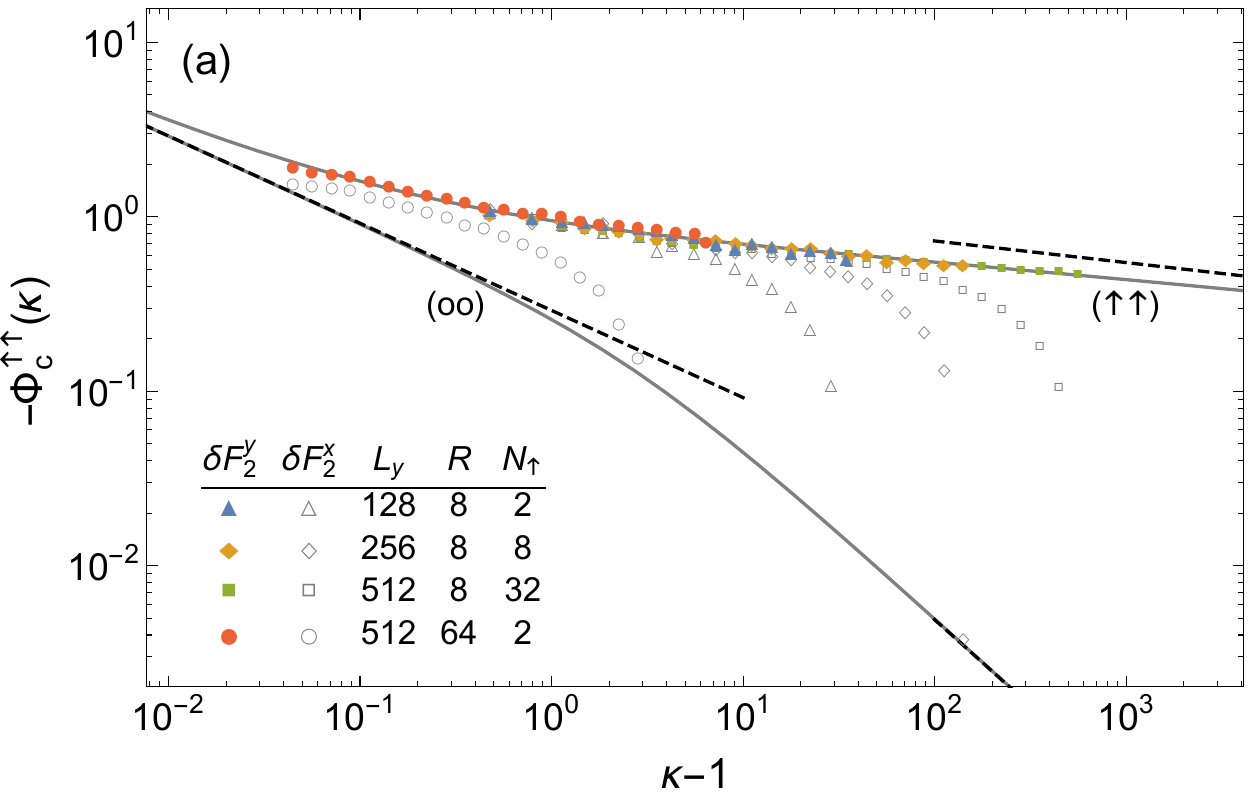}%
  	\hfill\hfill%
 	\includegraphics[width=0.48\textwidth]{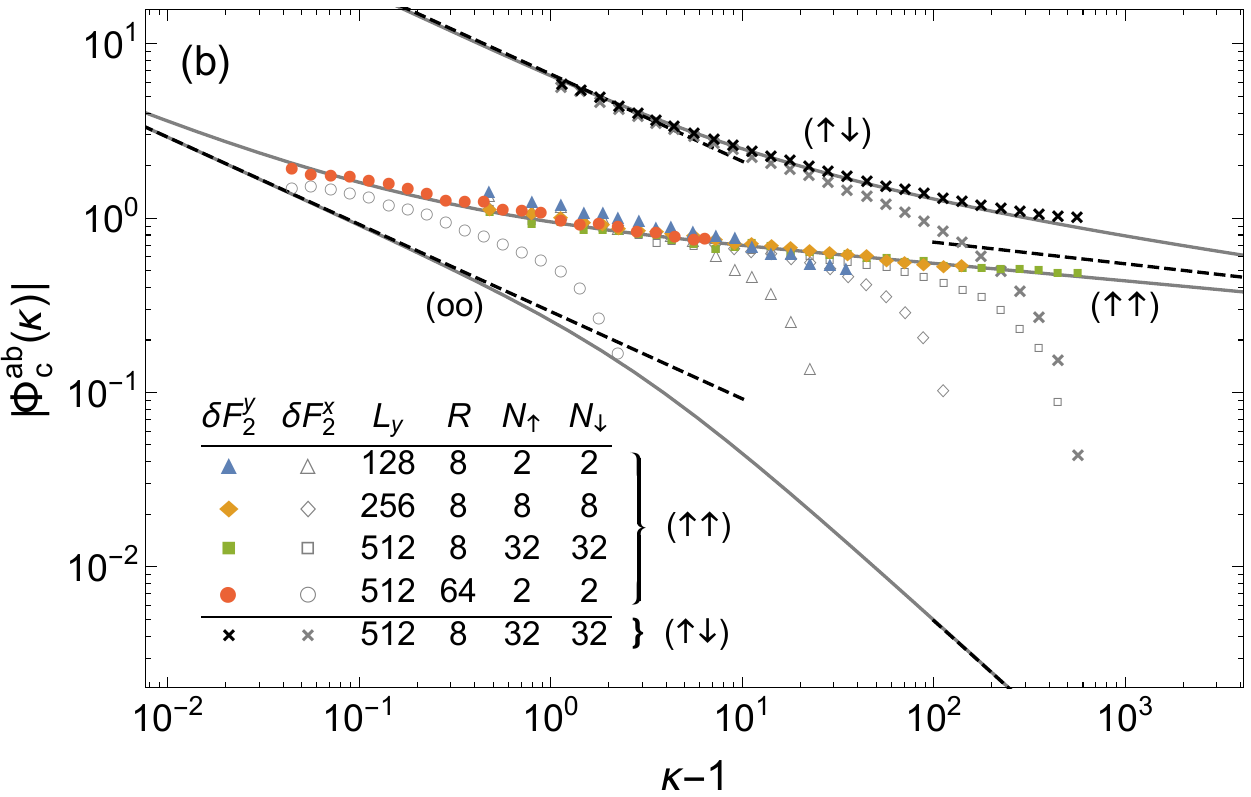}%
  	\hfill{}%
  	\caption{(Color online) Two-particle Casimir potential $\Fsc^{\uparrow\uparrow}(\kappa)$ in a system of size $L_{x}\times L_{y}$ at $T=\Tc$.
	The data points are $|\Fres_2(\vec r)|$ from the simulations, Eq.~\eqref{eq:ReversibleWork2}, while the solid lines are calculated using Eq.~\eqref{eq:exact}, and the dashed lines are the asymptotes from Eqs.~\eqref{eq:FscKappa} and \eqref{eq:asymp2}.
	(a) Simulation with fixed medium magnetization $M=0$ and $N_{\uparrow}$ $\uparrow$-spheres having radius $R$ using only the cluster algorithm introduced in the last chapter.
	The data nicely collapse onto the curve for the $\Fsc^{\uparrow\uparrow}$ scaling function.
	(b) Simulation with fluctuating medium magnetization due to an additional Wolff update.
	The symmetry of the medium is conserved by using the same number $N_{\uparrow}$ and $N_{\downarrow}$ $\uparrow$- and $\downarrow$-particles.
	Additionally the simulation results for the $\uparrow\downarrow$ interaction of the system with $L_{x}\times L_{y}=1024\times512$ and $N_{\uparrow}=N_{\downarrow}=32$ particles having radius $R=8$ is shown as black crosses.
	Smaller systems show larger deviations due to the repulsive interaction.
 }
 \label{fig:Fab2}
\end{figure*}

It is worth to comment on the deviations between our results and those of Machta \textit{et al.} \cite{Machta2012}:
It is crucial that the particles are large compared to the lattice spacings, i.e. $R\gg 1$, to guarantee to be in the correct limit for a comparison between the simulation results and the CFT predictions, since the theory requires the particles to be macroscopic objects in a continuous medium, coupling to many medium degrees of freedom.
We found that a radius of at least $R = 8$ is necessary to get the correct scaling behavior, see Fig.~\ref{fig:Fab1}.
If that is not the case, one only measures some kind of magnetization correlation function, which has the same large distance behavior but a different and non-universal prefactor.
Machta \textit{et al.} \cite{Machta2012} used too small disks with radius $R<4$ and compensated the deviations of their results by simply adding constants to each curve so that they match the CFT predictions at the farthest accessible simulation point. Additionally, with such small particles they cannot access the near field region $\kappa < 2$.

In order to resolve this discrepancy correctly, we needed to force the medium back into its critical state, i.e., restore its symmetry around $M=0$.
In the canonical ensemble this can be done by either fixing the magnetization to zero or by symmetrizing the polarization effect, while in the grand canonical approach as used in Ref.~\cite{ETBEvRD15} the mean magnetization could be fixed to zero by the chemical potential.
The first approach can be realized by using only the algorithm introduced in the last chapter and no Wolff update, see Fig.~\ref{fig:Fab2}a.
Since the algorithm conserves the order parameter, the system is fixed into a critical sub-ensemble (Fig.~\ref{fig:configurations}).
For the second approach we used the same number $N_{\uparrow}$ and $N_{\downarrow}$ of $\uparrow$- and $\downarrow$-particles, and a fluctuating magnetization, preserving the $Z_{2}$ symmetry of the system.
Additionally, this last approach allows us to simulate the Casimir potential for $\uparrow\uparrow$ and $\uparrow\downarrow$ boundary conditions \textit{simultaneously}, see Fig.~\ref{fig:Fab2}b.

Both approaches force at least two domains into the system, thus it turns out to be beneficial to change the systems aspect ratio to $1/2$, \ie, $L_x=2L_y$, as then two dominant domains fit into the system, see Figs.~\ref{fig:configurations}b and~\ref{fig:configurations}c.
Since the system tries to minimize the length of the domain walls, it is most likely to find the domains separated along the parallel direction.
Thus the system is highly anisotropic with respect to the parallel and the perpendicular directions, and we can measure the two-particle distribution function in the two directions independently.
We restrict our analysis of the two-particle distribution function to parallel and perpendicular strips with width $2R$ around the symmetry axes of the system centered in the middle of the exclusion volume.
We performed simulations with fixed order parameter in systems with $L_{y}\in\{128,256,512\}$ and $N_{\uparrow}\in\{2,8,32\}$ particles, respectively, all having radius $R=8$ and thus all with a particle volume fraction $\varrho\approx 0.012$.
This volume fraction seems to be small enough to suppress any many-body aggregation processes, as there are no evidence of clusters in the correlation function and thus in the free energy, too.
Additionally we performed a simulation with $L_{y}=512$ and two particles with radius $R=64$ to get values for small distances, i.e., $\kme\ll1$.
Additionally we performed those simulations again with fluctuating order parameter and $N_{\uparrow}=N_{\downarrow}$ particles, where again $N_{\uparrow}\in\{2,8,32\}$.
Note that the result for $R=64$ were corrected for excluded volume effects by dividing $g_{2}$ by $1-8\pi R^{2}/V$, a term which turns out to be negligible for $R=8$.
This factor accounts for the exclusion volume $\pi (2R)^{2}$ and the fact that the particles are restricted to about half the system due to the surface preference and the fixed magnetization $M=0$, see Figs.~\ref{fig:configurations}b and~\ref{fig:configurations}c.

For both approaches the resulting Casimir potentials $\Fres_2^{x,y}(\vec r)$ in \textit{both} directions show the expected scaling behavior at least at small distances and nicely agree with the exact scaling function $\Fsc^{\uparrow\uparrow}(\kappa)$ for equal symmetry breaking surface preferences.
For the perpendicular case, $\Fres_2^y(\vec r)$ perfectly agrees with the exact result up to the largest possible distances, while the Casimir potential $\Fres_2^x(\vec r)$ shows a drop-off for large $\kappa$, marked as gray symbols in Fig.~\ref{fig:Fab2}.
This is a consequence of the domain with opposite orientation caused by the conserved order parameter or the two different species of particles, respectively, which leads to an effectively repulsive force onto the particles.
The cutoff at $r\approx L_x/2$ follows from a change of sign in the potential due to this anti-correlation between the particles and the domain with opposite orientation.
The deviation from the predicted curve starts at about half this cutoff distance.

Figure~\ref{fig:Fab2}b additionally shows the results of $\Fres_2^y(\vec r)$ for $\uparrow\downarrow$ boundary conditions from the simulation with $L_{x}\times L_{y}=1024\times512$ and particles having radius $R=8$ as black crosses.
Again the data nicely collapses onto the exact scaling function $\Fsc^{\uparrow\downarrow}$.
For smaller systems with less particles the statistics is insufficient especially at small distances because of the repulsive character of the force and has thus stronger corrections for large distances due to the periodic boundary conditions.

We now head on to a special three-body problem that connects the two- and three-body interaction via a conformal mapping. 

\section{Conformal mapping for a special three-body configuration}

\begin{figure*}[t]
  \includegraphics[width=0.8\textwidth]{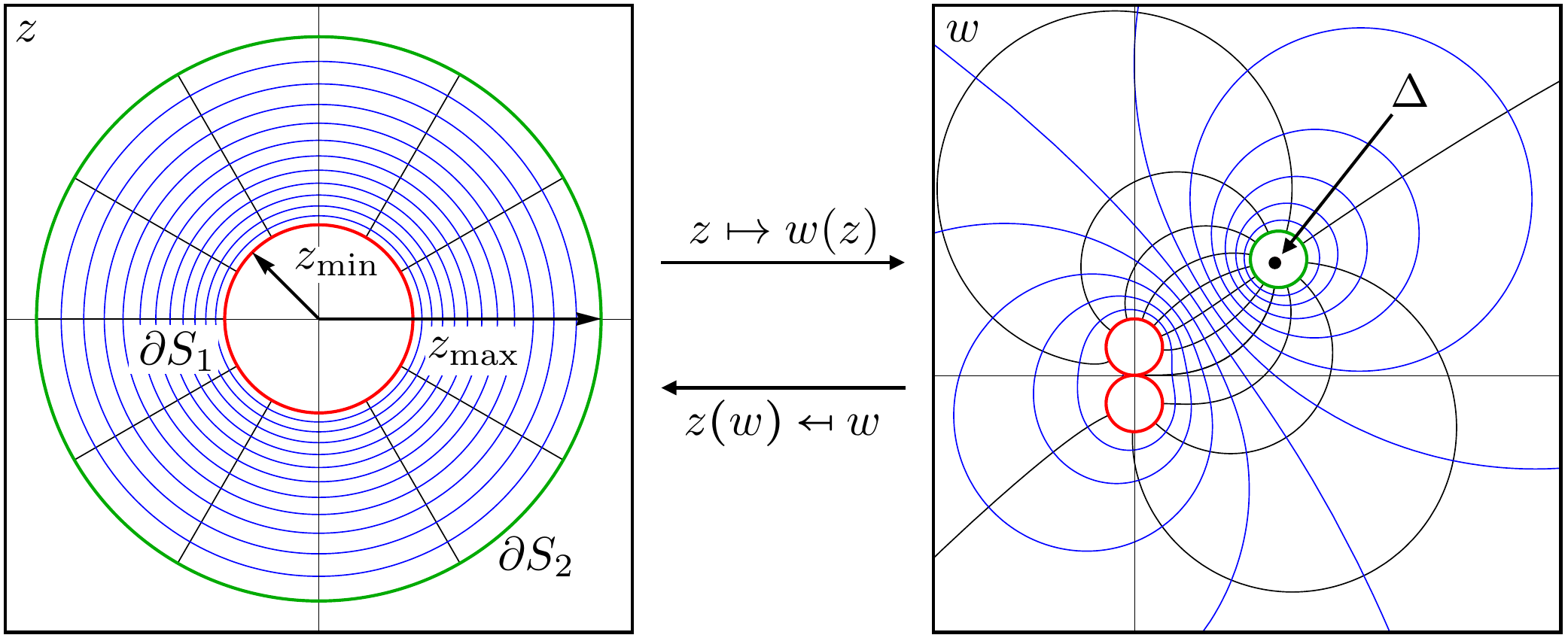}%
  \caption{(Color online)
  	Using Eq.~\eqref{eq:CFMapping}, the annulus is mapped onto the three-body setting with two disks in contact.
	The circle $\partial S_{1}$ with radius $\zmin=1$ is mapped onto two adjacent disks, while the outer circle $\partial S_{2}$ with radius $\zmax$, calculated with Eq.~\eqref{eq:rmax}, is mapped onto an object with approximately circular shape.
	Its position is given by the complex parameter $\Del$ and is centered at $\rdrei$, where $\rdrei\to\Del$ for $\zmax\to\infty$.
  }
  \label{fig:CFT_Map}
\end{figure*}

As we know the exact form of the two-body critical Casimir potential scaling function, we can use it to calculate a limiting case of a three-body interaction, where two particles touch and a third is allowed to move freely.
Since the first two particles are in contact, they can be understood as one deformed object and we can use a conformal mapping to transform the scaling function of the annulus geometry as proposed in \cite{BimonteEmigKardar13}.

To derive the mapping from the annulus to the setting of two adjacent disks and a third one free to move, we start with the M{\"o}bius transformation
\begin{align}
  z\mapsto \frac{\Delta z +\Delta^{\ast}}{z-1}
\end{align}
that maps the annulus with inner radius $\zmin=1$ onto the
half plane with $\mathrm{Re}(z)\geq 0$ and the point $z=\infty$ onto the point
$\Del\in\mathbb{C}$, where $\Del^{\ast}$ is the complex conjugate of $\Del$.
Taking the logarithm with an appropriate normalization factor maps the imaginary axes onto two parallels symmetric around the real axis.
An additional inversion maps them onto two symmetric circles with radius 1 touching at the origin, where a normalization factor $\pi$ is necessary.
To compensate for the logarithm, we map $\Del\mapsto \exp(\pi/\Del)$, so the point $z=\infty$ is again mapped onto the point $\Del$.
Thus the resulting conformal mapping reads
\begin{align}
  z\mapsto w(z)=\pi \left[\ln\left(\frac{z e^{\pi/\Del}+e^{\pi/\Del^{\ast}}}{z-1}\right)\right]^{-1}.
  \label{eq:CFMapping}
\end{align}
The inner radius of the annulus has to be chosen to be $\zmin=1+\epsilon$ with $\epsilon\to 0^+$ to ensure that the contour $\partial S_1$ of the two adjacent disks is a Jordan curve, while the outer radius $\zmax$ is by determined Green's theorem from the condition that the enclosed area fulfills
\begin{align}
  \frac{1}{2 i} \oint_{|z| = \zmax} w^\ast \mathrm d w = \pi.
  \label{eq:rmax}
\end{align}
The position $\rdrei\in\mathbb{C}$ of the third disk relative to the origin where the two adjacent disks touch is given as geometric center of its contour
\begin{align}
  \rdrei=\frac{1}{2\pi}\oint_{|z|=\zmax}w\, |\mathrm{d}w|.
  \label{eq:center}
\end{align}
Note that the last two integrals can only be performed numerically.
The third particle is not a perfect disk, but its shape varies as it comes very close to the other two disks, additionally depending on the phase of $\Del$.
For large $|\Del|$ its mean square deviation from a perfect unit circle becomes 
\begin{align}  \label{eq:r_3_error}
  \oint_{|z|=\zmax}(|w-\rdrei|-1)^2\, |\mathrm{d}w| = \mathcal{O}(|\Del|^{-8}),
\end{align}
giving a very good approximation even for small distances.

With this transformation we are able to calculate a special case of the three particle interaction from the two particle interaction, utilizing the transformation formula for the free energy scaling function from \cite{BimonteEmigKardar13}
\begin{align}
  \Phi_{w}^{ab}(\kappa')=\Fsc^{ab}(\kappa')-\frac{i}{12\pi}\int\limits^{\infty}_{\rdrei}\!\mathrm{d}\zeta\int\limits_{\partial S_{2}}\!\mathrm{d}w\,\{z,w\},
  \label{eq:CFTScaling}
\end{align}
where $\{z,w\}=(\partial^3_w z/\partial_w z)-(3/2)(\partial^2_w z/\partial_w z)^2$ is the Schwarzian derivative of the inverse mapping
\begin{align}
	w\mapsto z(w)=\frac{e^{\pi/w}+e^{\pi/\Del^{\ast}}}{e^{\pi/w}-e^{\pi/\Del}}
	\label{eq:InverseCFMapping}
\end{align}
and $\partial S_{2}$ is the mapped contour of the outer circle of the original annulus with radius $\zmax$.
The integral over $\zeta$ inserts the second object at the position $\rdrei$ relative to the center of the first one.
For the mapping Eq.~\eqref{eq:InverseCFMapping} the contour integral around $\partial S_{2}$ vanishes and thus the new scaling function is the original one with a modified scaling variable.
Since $\kappa$ only depends on the inner and outer radii of the original annulus, the new scaling variable reads
\begin{align}
  \kappa'=\frac{1}{2}\left(\frac{\zmax}{\zmin}+\frac{\zmin}{\zmax}\right)\approx\frac{\zmax+\zmax^{-1}}{2}.
\end{align}
In the far field limit we find the expansion
\begin{align} \label{eq:kappapExpansion}
  \kappa' = \frac{1}{\pi}|\rdrei|^2 - \left(\frac{1}{\pi}+\frac{\pi}{2}\sin^2(\arg \rdrei) \right) + \mathcal O(|\rdrei|^{-2}),
\end{align}
which gives a surprisingly good approximation for $\rdrei \gtrsim 3R$, see Fig.~\ref{fig:dF3rCFT}.
Comparing with Eq.~\eqref{eq:kappa} we conclude that to lowest order the two particles have the same far field as a single particle with effective radius $\hat R \equiv \frac{\pi}{2}R$. 
 This value is between the naive approximations $\hat R=\sqrt{2}R$ from fixed volume ($\pi \hat R^2=2\pi R^2$) and $\hat R=2R$ from fixed surface area ($2\pi \hat R=4\pi R$).
As our mapping becomes exact in the limit $|\rdrei|\to\infty$, we can calculate the exact far field amplitudes for several BC combinations in the considered geometry.

\section{Three-particle interaction}

If a third particle is getting close to two others, the pairwise description fails and three-body contributions become relevant because of the nonadditive character of the critical Casimir force.
In analogy to the two particle case, Eq.~\eqref{eq:ReversibleWork2}, this effect can be characterized by the according $n$-point distribution function $g_n(\vec r_1,\ldots,\vec r_n)$. 
The function $g_n$ is directly related to the $n$-particle Casimir potential $\Uctot_n$ via the reversible work theorem \cite{chandler1987introduction},
\begin{align}
  \label{eq:ReversibleWork}
  g_n(\vec r_1,\ldots,\vec r_n)
  = \mathrm e^{-\Uctot_n(\vec r_1,\ldots,\vec r_n)},
\end{align}
where $\Uctot_N(\vec r_1,\ldots,\vec r_\mu,\ldots,\vec r_N)$ is the change in the total free energy of a system with $N$ particles if particle $\mu$ is added to the system from infinite distances.

\begin{figure*}[t]
	\includegraphics[width=\textwidth]{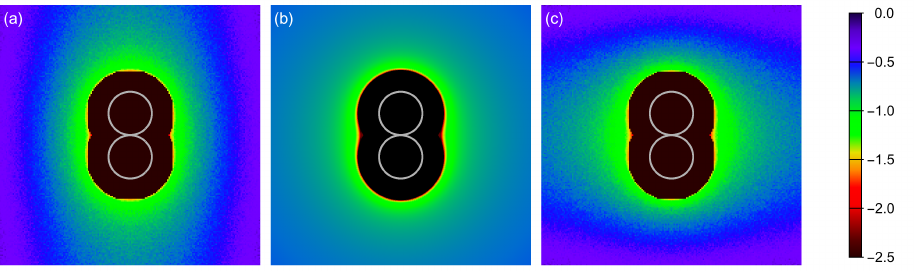}
	\caption{(Color online) Renormalized three-body Casimir potential $\Uctot_3^\mathrm{ren}(\vec{r}_{12},\vec{r}_{13})$ at $T=\Tc$.
	The first two particles (marked as gray circles) are fixed at their closest distance \fred{$r_{12}=2R$} as described in the text, while the third particle is allowed to move freely. 
	The exclusion volume is shown as black region.
	(a) and (c) show the potential for a system with $L_{x}\times L_{y}=512\times256$ and $N_\uparrow=3$ particles having radius $R=16$.
	\fred{The geometry of the whole system is shown as insets.}
	Note that (c) is rotated by $90^\circ$.
	(b) shows the corresponding infinite volume scaling function calculated with the conformal mapping, Eq.~\eqref{eq:CFMapping}.}
	\label{fig:UC}
\end{figure*}
\begin{figure*}[t]
	\includegraphics[width=\textwidth]{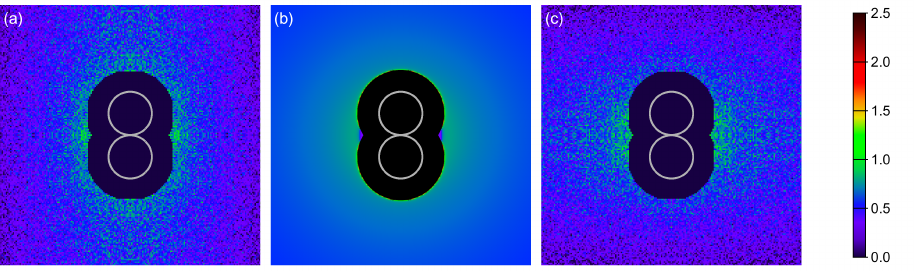}
	\caption{(Color online) 
	Same as Fig.~\ref{fig:UC}, but for the pure three-body contribution $\Uc_3(\vec{r}_{12},\vec{r}_{13})$, calculated with Eq.~\eqref{eq:Uc3}.
	$\Uc_3(\vec{r}_{12},\vec{r}_{13})$ is repulsive at short distances $r_{\mu 3}$ ($\mu=1,2$) but does not vanish as the distance between the third particle and the other two increases, as expected.
	Instead it has to compensate for the fact that $\Uctot_3^\mathrm{ren}(\vec{r}_{12},\vec{r}_{13})$ and $\Uc_2(\vec r_{\mu 3})$ are nearly equal for large $r_{\mu 3}$.}
	\label{fig:UCp}
\end{figure*}

Assuming that the influence of three or more particles is small compared to the two-particle interaction, it is common to decompose  $\Uctot_n$ into pure $k$-particles contributions $\Uc_k$ according to
\begin{align}
  \label{eq:Uctot}
\begin{split}
\Uctot_n(\vec r_1,\ldots,\vec r_n) &= 
 \sum_{\mu,\nu=1 \atop \mu<\nu}^n \Uc_2(\vec r_{\mu},\vec r_{\nu}) \\
&+ \sum_{\mu,\nu,\lambda=1 \atop \mu<\nu<\lambda}^n
\Uc_3(\vec r_{\mu},\vec r_{\nu},\vec r_{\lambda}) + \ldots\,.
\end{split}
\end{align}

We again assume translational invariance, with distance vectors $\vec r_{\mu\nu}=\vec r_\nu-\vec r_\mu$, and consequently all $n$-point functions depend on $n-1$ distances. 
For $n=2$ we get $\Uctot_2(\vec{r}_{12})=\Uc_2(\vec{r}_{12})$ and recover Eq.~\eqref{eq:ReversibleWork2}, while the three-particle Casimir potential decomposes to
\begin{align}
\label{eq:Uctot3}
\begin{split}
  \Uctot_3(\vec{r}_{12},\vec{r}_{13})={}&\Uc_2(\vec{r}_{12})+\Uc_2(\vec{r}_{13})+\Uc_2(\vec{r}_{23})\\
  {}+{}&\Uc_3(\vec{r}_{12},\vec{r}_{13}),
\end{split}
\end{align}
as it is used, e.g., in \cite{MattosHarnauDietrich2015}.
Therefore, the pure three-particle contribution can be calculated as
\begin{align}
\label{eq:Uc3}
  \Uc_3(\vec{r}_{12},\vec{r}_{13})=-\ln\!\left[\frac
  {g_3(\vec{r}_{12},\vec{r}_{13})}
  {g_2(\vec{r}_{12}) g_2(\vec{r}_{13}) g_2(\vec{r}_{23})}\right]
\end{align}
and requires the calculation of a four-dimensional histogram of the distances $g_3(\vec{r}_{12},\vec{r}_{13})$ in $d=2$ space dimension, which has to be accurately determined in the Monte Carlo simulations.
As such a histogram needs a lot of memory storage, it would limit our studies to small systems or to a lower resolution for the particle positions.
For a system with $L=256$ a naive approach would be to use an array with $(L/2)^{4}$ entries, which would require about $1\mathrm{GB}$ of memory storage.
Furthermore, to acquire a reasonable statistics we need, say, $100$ entries in each histogram bin on average, leading to $\approx 10^{10}$ independent measurements.
To considerably reduce the needed storage and simulation time we fixed the distance between the two particles 1 and 2, $\vec r_{12}=\mathit{const.}$, via a \textit{ghost bond} between them, \ie, an additional infinitely strong coupling between the spins at the center of each particle.
Note that it does not change the condition of detailed balance, because albeit the non-local coupling only the spins at the edge of the clusters contribute to the energy difference.
Those ghost bonds fix the relative position of two particles to each other and thus reduces the measurable distribution to one slice of the original histogram.
With this approach we are able to avoid a lot of configurations where the three-particle correlation is very small, \eg, when one particle is far away from the other two, which results in a better statistic and thus reduces the simulation time enormously.

We simulated a system with fixed order parameter at $M=0$ and $L_{x}=512$, $L_{y}=256$, $R=16$ and $N_{\uparrow}=3$ in the constellation with two particles adjacent at their closest approach coupled via a ghost bond, $\vec r_{12}=(2R,0)$ and $\vec r_{12}=(0,2R)$, and the third particle free to move independently.
Note that in the continuum limit both $\Uctot_3(\vec{r}_{12},\vec{r}_{13})$ and $\Uc_2(\vec{r}_{12})$ diverge if $r_{12} \to 2R$, while the difference 
\begin{align} \label{eq:dF3_ren}
\Uctot_3^\mathrm{ren}(\vec{r}_{12},\vec{r}_{13})\equiv\Uctot_3(\vec{r}_{12},\vec{r}_{13})-\Uc_2(\vec{r}_{12})
\end{align}
remains finite.
Figure~\ref{fig:UC} shows the resulting renormalized interaction potentials $\Uctot_3^\mathrm{ren}(\vec{r}_{12},\vec{r}_{13})$, while the pure three-body contributions $\Uc_3(\vec{r}_{12},\vec{r}_{13})$ according to Eq.~\eqref{eq:Uc3} are shown in Fig.~\ref{fig:UCp}.
The left plot shows a system where the first two particles are aligned in $y$ direction while for the right one they are aligned along the $x$ direction.
The middle frame shows the according scaling function calculated from the conformal mapping Eq.~\eqref{eq:CFTScaling}.
Contrary to the expected behavior, the pure three-body potential $\Uc_3(\vec{r}_{12},\vec{r}_{13})$ does not vanish if the third particle moves away from the other two, but decays very slowly with $\rdrei^{-1/8}$ just like the two-body interaction potential.
The three-body interaction is not the sum of the three two-body interactions but rather the sum of approximately two of them.
Thus the pure three-body contribution has to compensate for this overestimation, at least for large distances.
For the near field it seems that the divergences of the three-body interaction $\Uctot_3$ and the corresponding two-body terms in Eq.~\eqref{eq:Uctot3} cancel each other and that $\Uc_3$ remains finite as the third particle gets close to the other two.
The approximation with the conformal mapping fails as the particles form a triangular constellation for $|\rdrei|\lesssim 3 R$, since then the deformation of the third particle is no longer negligible.
This can be seen in Fig.~\ref{fig:UCp}b, where $\Uc_3$ is getting smaller, although the simulations predict an almost constant behavior.
In the far field we find $\Uc_3 \approx -\Uctot_3^\mathrm{ren} \approx -\Uc_2 $, see Figs.~\ref{fig:UC} and \ref{fig:UCp}, leading to a repulsive contribution to the three-body Casimir force.
This effect is stronger than the mean-field results in \cite{MattosHarnauDietrich2015} due to the stronger fluctuations in two dimensions.

\begin{figure}[t]
	\includegraphics[width=0.48\textwidth]{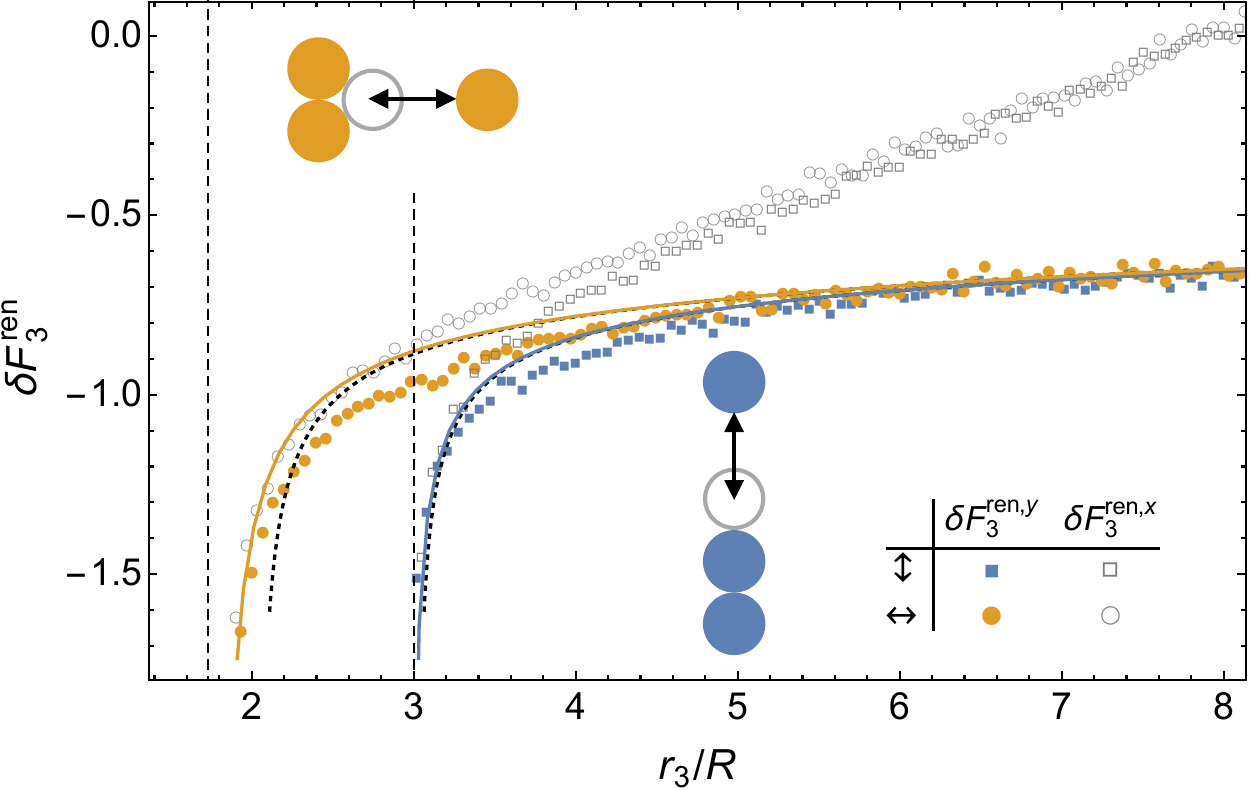}%
	\caption{(Color online) Comparison between the conformal mapping (lines), Eq.~\eqref{eq:CFTScaling}, and the simulation results (symbols) for the three-body interaction $\Uctot_3^\mathrm{ren}$, Eq.~\eqref{eq:dF3_ren}.
	The blue (dark) line marks the interaction strength along a linear configuration of the three disk, while the orange (light) line is along the other symmetry axis.
	The symbols match with the curves from the mapping after a correction with an effective radius $\Reff$ with $\delta R=-0.8(1)$.
	The dotted curves are calculated using Eq.~\eqref{eq:kappapExpansion}.
  	 }
  \label{fig:dF3rCFT}
\end{figure}

We analyzed the data shown in Fig.~\ref{fig:UC} for strips of width $2R$ along the two symmetry axes as done for the two-particle case, and found them to be in good agreement with the scaling function calculated with the conformal mapping after correcting the radius of the disks again with $\Reff=R+\delta R$ and $\delta R=-0.8(1)$. 
The results are shown in Fig.~\ref{fig:dF3rCFT}.
As seen for the two-particle interaction, the three-body interaction deviates from the theoretical curve if measured along the $x$ direction but fits nicely along the $y$ direction due to the periodic boundary conditions and the domain with opposed orientation. 

Our results are similar to the experimental measurement of three-body interactions by Brunner \etal \cite{BrunnerDobnikar04}; they studied a system of three charged colloids with a repulsive electrostatic interaction.
Although the three-body interaction was repulsive, too, the pure three-body interaction was attractive and thus showed just the opposite behavior as the two-body interactions.
But the maximum of the pure three-body interaction $\Uc_3$ is -- contrary to the previous assumption -- of the same order as the two-particle interaction and thus the assumption of a small and fast decaying three-body contribution to the total potential made for the decomposition in Eq.~\eqref{eq:Uctot} does not hold true anymore.

Basically both the MC results and the conformal mapping state that the number of interacting particles has only a very small influence on the interaction strength, but rather the pure existence of a surface in two different areas leads to the critical Casimir force.
The non-additivity of the Casimir interaction thus has some interesting consequences for the behavior of clusters of those particles.
If we consider two near-spherical clusters of colloidal particles with particle numbers $n_1$ and $n_2$ in $d$ dimensions and utilize Eqs.~\eqref{eq:kappa} and \eqref{eq:asymp2}, the Casimir interaction at large distances $r$ scales as
\begin{align}
\delta F_{n_1,n_2} (r) \sim - (n_1 n_2)^{\expo_\psi} r^{-2 x_\psi},
\label{eq:dFCluster}
\end{align}
with scaling exponent $\expo_\psi = 2 x_\psi / d$.
The interaction is asymptotically additive at large distances only for $\expo_\psi = 1$, while for $\expo_\psi < 1$ ($\expo_\psi > 1$) we find subadditive (superadditive) Casimir interactions, respectively. 
For the two cases of symmetry breaking ($\uparrow\uparrow$ and $\uparrow\downarrow$) and symmetry preserving ($\uparrow\!\!\ord$, $\downarrow\!\!\ord$, $\ord\ord$) BCs we find from well known exponent relations:
{
\setlength{\leftmargini}{12pt}
\setlength{\leftmarginii}{20pt}
\begin{itemize}\itemsep2pt \parskip0pt \parsep0pt
\item
Symmetry breaking BCs always lead to subadditive interactions, because the exponent $\expo_\sigma = 1-\gamma/(d\nu)$, and the susceptibility exponent $\gamma$ is always positive.
\item
With symmetry preserving BCs we get $\expo_\epsilon = 1-\alpha/(d\nu)$, and the condition for additivity in Eq.~\eqref{eq:dFCluster} reduces to $\alpha = 0$, while for $\alpha > 0$ ($\alpha < 0$) we find subadditive (superadditive) Casimir interactions, respectively. Therefore, we predict weak superadditive Casimir interactions for symmetry preserving colloids in superfluid $^4$He, which is in the three dimensional XY universality class, with $\expo_\epsilon \approx 1.007$.
\end{itemize}
}
\noindent An overview of exponent values for common $O(n)$ universality classes are given in Table~\ref{tab:exponents}.

\begin{table}[t]
\caption{Scaling exponent $\expo_\psi$ for common $O(n)$ universality classes in $d$ dimensions. 
Magnetic exponents relevant for symmetry-breaking boundary conditions carry the index $\sigma$, while energetic exponents relevant for symmetric BCs have the index $\epsilon$.
If $\expo_\psi<1$ the Casimir interaction is subadditive, while for $\expo_\psi>1$ the interaction is superadditive.
\label{tab:exponents}}
\begin{ruledtabular}
\begin{tabular}{cccccc}
$d$ & $n$  & $x_\sigma$ & $x_\epsilon$ & $\expo_\sigma$ & $\expo_\epsilon$ \\
\colrule
$2$ & $1$ & $ 1/8 $ & $ 1 $ & $ 1/8 $ & $ 1 $ \\
\colrule
$3$ & $1$ & $ 0.518151(6) $\footnotemark[1] & $ 1.41264(6) $\footnotemark[1] & $ 0.345434(4) $ & $ 0.94176(4) $ \\
$3$ & $2$ & $ 0.51905(10) $\footnotemark[2] & $ 1.51124(22) $\footnotemark[2] & $ 0.34603(7) $ & $ 1.00749(15) $ \\
$3$ & $3$ & $ 0.51875(25) $\footnotemark[3] & $ 1.5939(10) $\footnotemark[3] & $ 0.34583(16) $ & $ 1.0626(7) $ \\
$3$ & $\infty$ & $ 1/2 $ & $ 2 $ & $ 1/3 $ & $ 4/3 $ \\
\colrule
$4$ & $ $ & $ 1 $ & $ 2 $ & $ 1/2 $ & $ 1 $ \\
\end{tabular}
\end{ruledtabular}
\footnotetext[1]{Taken from Ref.~\cite{Simmons-Duffin15}.}
\footnotetext[2]{Taken from Ref.~\cite{CHPV06}.}
\footnotetext[3]{Taken from Ref.~\cite{CHPRV02}.}
\end{table}

\section{Conclusions}

We presented a highly efficient cluster MC algorithm for the simulation of colloids immersed in a binary liquid, based on the geometric cluster algorithm by Heringa and Bl{\"o}te \cite{HeringaBl98}.
The algorithm suppresses the effects of critical slowing-down near criticality at least at sufficiently low particle densities.
It can be extended to contain additional interactions between the particles, such as electrostatic forces as present in experiments.
We used this algorithm to calculate the critical two-particle Casimir potential $\Fres_2(\vec r)$ over a range in the distance $\vec r$ that governs four orders of magnitude in the according conformal invariant scaling variable $\kappa$.
We found a strong dependency on whether the symmetry of the medium order parameter is conserved or broken.
In the latter case our MC results differ from the expected scaling function $\Fsc^{\uparrow\uparrow}$, and instead deviates towards the scaling function for open boundaries with growing particle volume fraction $\varrho$.
In the former case the simulation agrees excellently with the scaling function $\Fsc^{\uparrow\uparrow}(\kappa)$ predicted for equal symmetry-breaking boundary conditions \cite{BurkhardtEisenriegler95}.
The deviating behavior can be understood as a finite-size effect and is related to the strong polarization of the medium in periodic systems, which shifts the system away from criticality. 
We could show that this effect can be suppressed by either using a fixed magnetization $M=0$ or inserting the same amount of particles with opposite surface preferences. 
The inevitable domain structure could be controlled by simulating rectangular systems with an aspect ratio $1/2$.

Finally we presented first results for the three-body Casimir potential $\Uctot_3(\vec{r}_{12},\vec{r}_{13})$. 
We could significantly speed up the determination of the required three particle correlation function $g_3(\vec{r}_{12},\vec{r}_{13})$ by introducing a ghost bond between particles 1 and 2, fixing their distance vector $\vec r_{12}$ to a fixed value.
The results show the same qualitative behavior as experiments on similar systems \cite{BrunnerDobnikar04}, but the pure three-particle contribution violates the assumption for the decomposition approach, i.e., it is not small compared to the two-particle interaction, which gives a strong hint that such an expansion does not converge.
This is confirmed by the calculation for the case of two adjacent disks interacting with a third one using a conformal mapping of the known case of the annulus geometry.
For this setup we find that the Casimir interaction between the two disks in contact and the third disk is almost identical to the interaction between two separated disks, emphasizing the non-additivity of the Casimir potential.
We quantified this non-additivity by introducing a scaling exponent $\expo_\psi$ which characterized the far-field behavior of two interacting particle clusters.
For the common universality classes we find both subadditive as well as weak superadditive Casimir interactions.

In three dimensions the algorithm may be applied to systems investigated experimentally like thin quasi two-dimensional films or clustering effects in a fluctuating bulk near criticality.
Therefor an additional chemical potential can be applied to the medium to simulate a grand canonical ensemble as done in \cite{ETBEvRD15} and investigate medium-magnetization-induced phase transitions.
Additional forces between the particles may give a more realistic setting, e.g., one could add Coulomb forces to simulate the electrostatic character of the silicon particles commonly used in experiments.
With higher volume fractions $\varrho$ a density-induced clustering process and demixing transitions may be observable \cite{SoykaZvyaHertHeldBech08}.

Since our algorithm allows for changes in the temperature, it is possible to calculate the temperature dependency of the scaling function and thus study the corrections to conformal field theory away from criticality.
Finally, the algorithm can be extended to include Janus or patchy particles with inhomogeneous surface preferences, which show interesting agglomeration behavior \cite{DrZvyagolskaya_en}.

\begin{acknowledgments}
We wish to thank F.~M.~Schmidt, M.~Hasenbusch, H.~W.~Diehl and R.~Evans for fruitful discussions.
This work was supported by the Deutsche Forschungsgemeinschaft through Grant No.~HU 2303/1-1.
\end{acknowledgments}

\cleardoublepage
\bibliography{Physik}

\end{document}